# Retrospective, Observational Studies for Estimating Vaccine Effects on the Secondary Attack Rate of SARS-CoV-2

Marlena Bannick[1], Fei Gao[1,2,3], Elizabeth Brown[1,2,3], Holly Janes[1,2,3]

*1: Department of Biostatistics, University of Washington*

*2: Vaccine and Infectious Diseases Division, Fred Hutchinson Cancer Center*

*3: Public Health Sciences Division, Fred Hutchinson Cancer Center*

*Corresponding Author*

Marlena Bannick

mnorwood@uw.edu

## Abstract

COVID-19 vaccines are highly efficacious at preventing symptomatic infection, severe disease, and death. Most of the evidence that COVID-19 vaccines also reduce transmission of SARS-CoV-2 is based on retrospective, observational studies. Specifically, an increasing number of studies are evaluating vaccine efficacy against the secondary attack rate of SARS-CoV-2 using data available in existing healthcare databases or contact tracing databases. Since these types of databases were designed for clinical diagnosis or management of COVID-19, they are limited in their ability to provide accurate information on infection, infection timing, and transmission events. In this manuscript, we highlight challenges with using existing databases to identify transmission units and confirm potential SARS-CoV-2 transmission events. We discuss the impact of common diagnostic testing strategies including event-prompted and infrequent testing and illustrate their potential biases in estimating vaccine efficacy against the secondary attack rate of SARS-CoV-2. We articulate the need for prospective observational studies of vaccine efficacy against the SARS-CoV-2 SAR, and we provide design and reporting considerations for studies using retrospective databases.

# 1 Introduction

COVID-19 vaccines were developed primarily to prevent symptomatic infection and most importantly severe disease and death, and have shown high efficacy against these endpoints in experimental and observational studies.[1–13] There is evidence that these vaccines also prevent infection [5,14–18] and potentially reduce transmission,[19–23] albeit with smaller effects against the highly transmissible Omicron variant as compared to wildtype SARS-CoV-2 and earlier variants.[24–26]

While randomized controlled trials have played a key role in our understanding of efficacy against COVID-19 disease, they are inefficient for capturing asymptomatic SARS-CoV-2 infection or transmission and so most evidence for COVID-19 vaccine effects against these endpoints comes from observational studies. Recent literature has evaluated study designs and analytical approaches for estimating vaccine efficacy against infection.[27–30] Here, we focus on another key component of a vaccine's effect on the overall burden of the epidemic: its ability to reduce secondary transmission.[30–33]

The secondary attack rate (SAR) – the rate of transmission to susceptible contacts within a transmission unit over a well-defined time period [34] – is a common measure of transmission risk. Vaccine efficacy against the SAR is one minus the relative SAR for infected vaccinated individuals versus the SAR for infected un-vaccinated individuals.[31,34,35] Evaluating vaccine efficacy against the SAR of SARS-CoV-2 is challenging in the experimental setting for several reasons, particularly in the context of an ongoing pandemic.[28,36] It requires frequent surveillance of trial participants to detect all incident infections, even those asymptomatic, and enrollment and surveillance of all contacts of infected participants to detect secondary transmission. Phase III COVID-19 vaccine efficacy trials generally captured asymptomatic infections only through periodic SARS-CoV-2 serology, and infectiousness was only assessed through the proxy of SARS-CoV-2 viral load in symptomatic individuals; transmission events from infected trial participants were not captured.[2,3,5–8,33,37–40] Phase III efficacy trials also provide limited evidence on newly emerged variants of concern, e.g., the Delta and Omicron variants, because the blinded follow-up was largely complete before their emergence.

Given these challenges, most data on COVID-19 vaccine efficacy against the SARS-CoV-2 SAR come from retrospective observational, post-licensure studies, where systematic diagnostic testing is often absent.[33] In particular, studies of vaccine efficacy against the SARS-CoV-2 SAR are often done using linked national registries of vaccination and infection data based on household addresses,[21–23] and of contact tracing databases.[19,20] In this manuscript, we illustrate the limitations of retrospective, observational study designs for studying vaccine efficacy against the SARS-CoV-2 SAR, beyond the limitation of all observational studies which is potential confounding. First, we highlight the main classes of study designs. Next, we discuss challenges in the ascertainment of infections and transmission events using these designs. We derive an analytical formula for the bias in vaccine efficacy against the SAR introduced by common diagnostic testing strategies in these designs. Finally, we articulate the need for well-designed prospective observational studies of vaccine efficacy against the SARS-CoV-2 SAR, provide study design and reporting considerations for retrospective designs, and highlight areas for methodological innovation that may enable more accurate estimation of vaccine efficacy of the SARS-CoV-2 SAR when retrospective databases are the only option.

## 2 Retrospective study designs for evaluating VE against the SARS-CoV-2 SAR

There are two primary retrospective designs used by existing studies that evaluate VE against the SARS-CoV-2 SAR: record-linked database studies, and contact-tracing database studies. We discuss the features of these two study designs with a focus on identifying transmission units and transmission events (see Table 1 for examples of study designs).

### 2.1 Identifying transmission units in retrospective designs

Key to studying SARS-CoV-2 transmission is identifying the cohort of interest: susceptible contacts of infected individuals at high risk of SARS-CoV-2 due to exposure to the infected individual. The group of individuals that includes an infectious person and their susceptible contacts is commonly referred to as a *transmission unit.*[34] We use the term *index case* for the first detected case and *primary case* for the first infected person in the transmission unit.[34,41] Transmission units may be identified based on a characteristic at the group level, such as a shared residence, or they may be defined in reference to an individual, e.g., close contacts identified through contact tracing. Defining a transmission unit requires defining "susceptible," which may account for prior immunity due to vaccination and/or prior infection. Given data suggesting that breakthrough SARS-CoV-2 infections [42–44] and re-infections [45–48] are not uncommon, especially in the context of emerging variants, there are arguments for considering all contacts to be susceptible.

#### *2.1.1 Health care record linkage to infer transmission units*

An important class of retrospective designs defines a household as the transmission unit and identifies SARS-COV-2 transmission events within households by leveraging databases of diagnostic test results; vaccination status, type, and timing; and household addresses. For example, Prunas and colleagues [22] used data from household members in a database from a health maintenance organization covering 2.5 million individuals in Israel that includes demographics, healthcare utilization, vaccination, and SARS-CoV-2 test results. Lyngse and colleagues [23] used unique identification numbers for residents of Denmark to link reverse transcription polymerase chain reaction (PCR) and antigen test results from the Danish Microbiology Database with vaccination records in the Danish Vaccination Register and identified transmission units based on residential address. Similarly, Harris and colleagues [21] linked the national immunization database in England to a dataset with all laboratory-confirmed SARS-CoV-2 infections and identified persons sharing the same address. Transmission units captured via record linkage may incorporate some degree of measurement error due to changes in households over the course of the study or non-traditional household structures.[49,50]

### *2.1.2 Contact tracing databases to infer transmission units*

An alternative retrospective design makes use of existing SARS-CoV-2 contact tracing databases that include diagnostic test results of contacts of confirmed cases; the databases are linked with databases of vaccination records. One example is Gier and colleagues,[20] where close contacts of confirmed COVID-19 cases were sought out for PCR testing by the Municipal Health Services contact monitoring in the Netherlands, and the test results were combined with the national infectious disease notification registry that contains vaccination records. Similarly, Eyre and colleagues [19] used data from England's National Health Service Test and Trace Service, which performs a similar contact monitoring service for individuals who test positive in England. Vaccination status was obtained from National Immunization Management Service. Importantly, transmission units inferred from contact tracing databases are specific to the definitions, strategies, and function of the tracing service. Transmission units may be incompletely captured if tracing services are overly conservative in defining contacts of infected individuals, if infected individuals are reticent to disclose their contacts, or if contacts cannot be contacted or tested; incomplete capture of contacts is a pervasive challenge for contact tracing systems.[51–53] Mutually exclusive transmission units, which are ideal for direct estimation of the SAR, may be difficult to define using contact tracing databases, where "strings" of infections and contacts are identified.

## 3 Impact of testing strategy in retrospective designs

Retrospective designs, by definition, rely on testing designed for clinical diagnosis or management as opposed to research. One common challenge in observational settings is 'event-prompted' SARS-CoV-2 diagnostic testing, i.e., testing triggered by occurrence of an event, such as onset of symptoms, potential SARS-CoV-2 exposure, or testing mandated by social institutions or circumstances, e.g., travel, employment, or medical procedures.[33] Another challenge is infrequent testing.[29] In this section, we describe the potential pitfalls of event-prompted and infrequent diagnostic testing and illustrate these with a quantitative analysis of potential biases associated with these testing strategies.

## 3.1 Limitations of event-prompted testing

The most common SARS-CoV-2 test trigger is presentation of symptoms. However, symptom-prompted testing is guaranteed to miss many SARS-CoV-2 infections, given the large burden of asymptomatic infection.[54,55] The proportion of infections that are asymptomatic varies widely and is associated with individual-level factors such as age, comorbidities, pre-existing immunity;[54,56] method for ascertaining symptoms;[57] and infecting variant.[58] The extent to which existing databases capture primarily symptomatic testing may vary over time and geography as routine asymptomatic testing of individuals becomes more common or is mandated by social institutions. Many studies have no information on what prompted the tests captured in the database, which prevents any deliberation or exploration as to the impact of the missed infections on vaccine efficacy estimates.

In addition to missed asymptomatic infections, symptom-prompted testing poses challenges in inferring transmission chains because infections are not detected until (and if) symptoms occur, typically several days after acquisition of infection. The incubation period for SARS-CoV-2 — the time between onset of infection and symptoms — has been estimated as 6 days on average for ancestral strains and shorter with new variants, and is variable across individuals.[59–62]

## 3.2 Limitations of infrequent testing

Even if testing is routine, infections may still be missed if tests are performed infrequently. The duration of SARS-CoV-2 positivity by diagnostic testing is variable and is as short as 1 day for some individuals,[63–65] especially when vaccinated or previously infected. And yet, even individuals who shed for a short period may still have transmission potential if shedding large amounts of virus.[66] Daily PCR testing is likely necessary to capture all SARS-CoV-2 infections, and twice-weekly PCR testing likely captures a supermajority of infections.[67–70]

SARS-CoV-2 serology is another potential tool for capturing current or past infections based on periodic blood collection. While commercial assays detecting antibodies against the nucleocapsid protein (anti-N antibodies) have demonstrated high sensitivity and specificity,[71] and anti-N antibody responses are reported to be durable [72] and are not elicited by COVID-19 vaccines that target the spike protein, serology has limited value for studies of transmission because it does not inform on timing of infection. Furthermore, serological assays may be insensitive to past infection in vaccinated individuals.[73]

## 3.3 Potential impact of event-prompted or infrequent testing on estimation of vaccine efficacy

To illustrate the potential impact of event-prompted or infrequent testing on estimation of VE against transmission, we compare the "target" estimand — the expectation of the statistical estimate of vaccine efficacy against the SARS-CoV-2 SAR — with the "actual" estimands under imperfect testing. We assume a range of scenarios around SARS-CoV-2 infection and transmission (see Appendix A) that are intended to illustrate the *issues of symptom-prompted testing and infrequent testing in isolation*; in practice there may be several issues at play.

Under the scenarios we consider, for both symptom-prompted testing and infrequent testing the actual estimand is smaller than the target estimand, except when vaccinated individuals have no transmission potential in which case the target and actual estimands agree. As the transmission potential of breakthrough infections increases, the difference between the estimands increases. Using symptom-prompted testing, this relationship is more pronounced for larger reductions in transmission for asymptomatic infections (Figure 1[a]). When asymptomatic individuals have the same transmission potential as symptomatic individuals the estimands agree regardless of VE against the SAR. The implication is that, since asymptomatic individuals likely transmit less frequently than symptomatic individuals,[74] testing only symptomatic infections may tend to underestimate vaccine efficacy against the SARS-CoV-2 SAR.

Under infrequent testing, the difference between estimands is a function of the testing interval: whenever the testing interval is longer than the minimum duration of infection (1 day), the actual estimand is smaller than the target estimand (Figure 1[b]). In Figure 1[b], we show only testing intervals that are shorter than the maximum duration of infection in the vaccinated group (15 days). Generally, the relationship between the testing interval and the difference between the actual and target estimands is not monotonic, and when the interval is longer than the longest duration of infection, the difference between the estimands does not depend on the testing interval. The implication is that infrequent testing would tend to yield an underestimate of vaccine efficacy against the SARS-CoV-2 SAR.

***Figure 1.*** *Analytical comparison of actual (y-axis) and target (x-axis) vaccine efficacy against SARS-CoV-2 transmission estimands. The dashed line indicates equality. (a) Difference between target and actual estimands under symptom-prompted testing under varying reduction in transmission potential for asymptomatic versus symptomatic infections; (b) Difference between target and actual estimands under infrequent testing under varied frequency of testing. Note the different scales of the two y axes. In (a), the x-axis is the one minus the ratio of the secondary attack rate comparing asymptomatic to symptomatic infected people: a value of 1.0 indicates that asymptomatic infected people cannot transmit SARS-CoV-2, and a value of 0.0 indicates that the transmission potential for asymptomatic and symptomatic people is the same. In (b), the x-axis is the interval between subsequent tests, e.g., a value of 7 means that individuals were tested once every 7 days.*

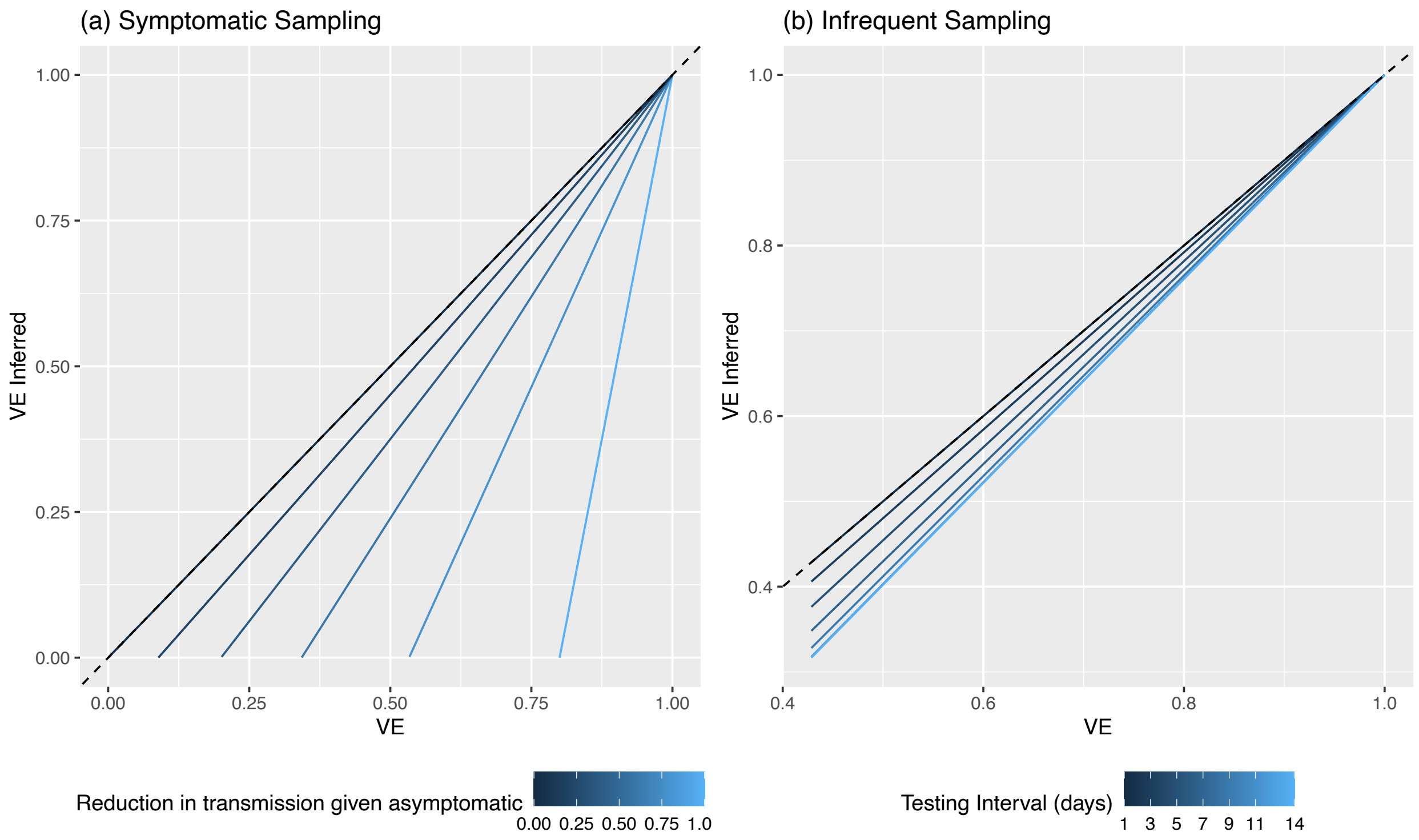

In Figure 1, we assumed correct classification of who infected whom. Figure 2 illustrates how irregular and infrequent testing for SARS-CoV-2 may also misclassify transmission events. In the hypothetical transmission units shown (Figure 2[a]), both symptom-prompted testing (Figure 2[b]) and infrequent testing (Figure 2[c]), underestimate the number of transmission events and result in misclassifying the primary case in at least one of the transmission units. Under symptom-prompted testing (Figure 2[b]), the misclassification is due to a longer pre-symptomatic period for the primary case vs. the secondary infection. The impact of misclassification of transmission events will depend on the extent and nature of the misclassification; the actual estimand may be higher or lower than the target estimand.

*__Figure 2.__ True and inferred transmission chains for two hypothetical transmission units, each with one primary case and three susceptible contacts. Black dots represent transmission events [in panels (a) true events, (b-e) inferred transmission events]. Panels (b-c) show inferred transmission events using symptom-prompted or infrequent testing strategies; panels (d-e) show inferred transmission events when contact-to-contact or community-acquired infections are incorrectly included as transmission events.*

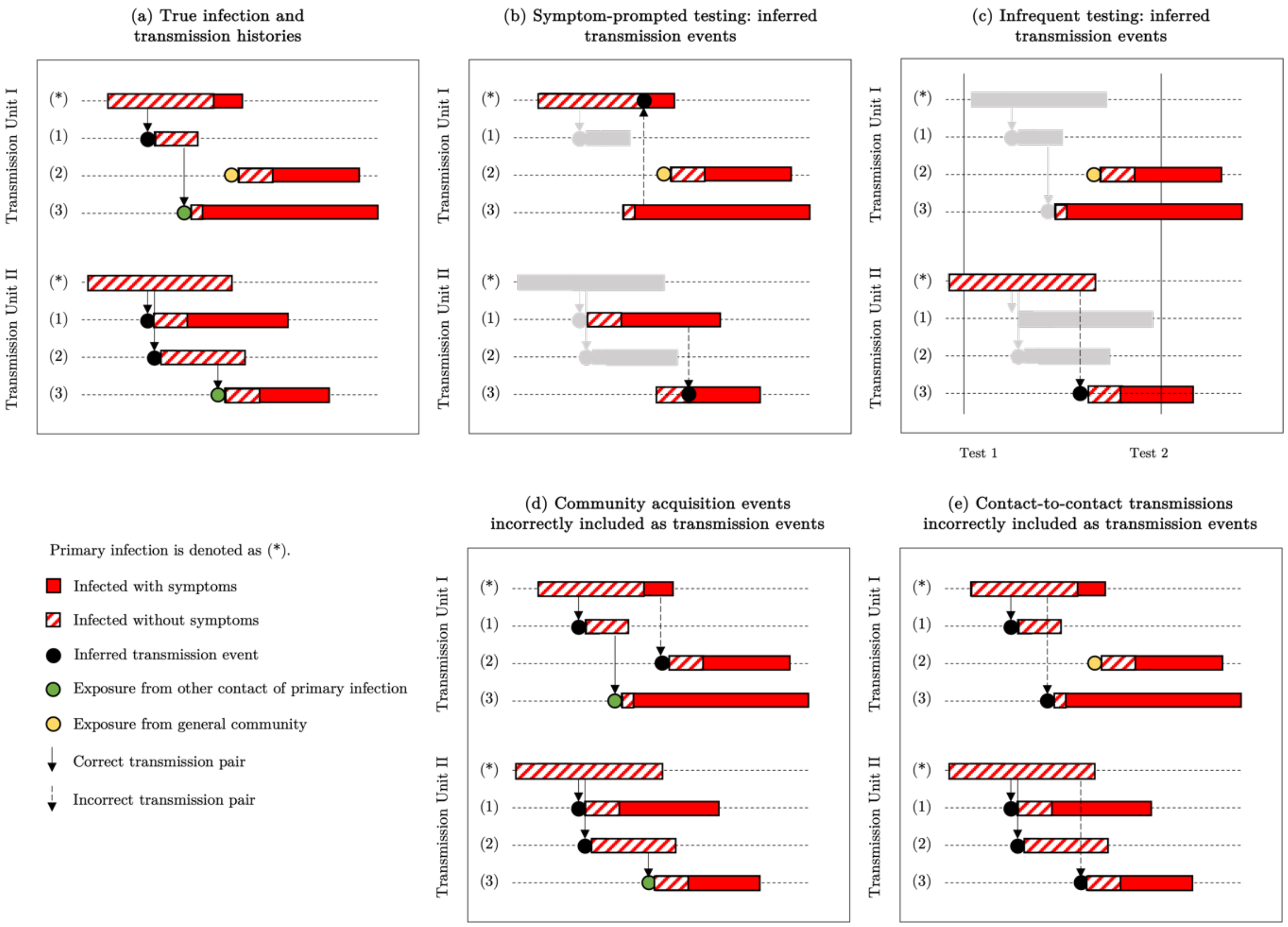

## 3.4 Impact of incomplete testing of transmission unit

Another often-overlooked challenge with observational studies is that some members of a transmission unit may not be tested for reasons beyond event-prompted or infrequent testing. For example, some individuals may 'opt out' of testing entirely or during certain time periods, or there may be limited testing availability. Such incomplete testing is likely nearly universal and is not always reported. Two out of three of the household studies listed in Table 1 did not report testing rates; Lyngse and colleagues [23] reported that about 10% of household members in their study were not tested within one week after the index case tested positive. For records-based studies, the most common approach is to assume that individuals not tested are SARS-CoV-2 negative. However, the outcome for untested individuals is missing. The result is that there will be incomplete capture of both primary case and secondary transmission events. For contract tracing studies, the number of susceptible contacts — the denominator of the SAR — may also be underestimated due to incomplete testing. Often it is difficult to know whether missingness is 'completely at random,'[75] or whether missed infection events depend on factors such as vaccination or demographics, and the direction of the bias that is induced. We revisit the issue when discussing data analytic strategies.

## 3.5 Strategies for confirming potential transmission events

Correct estimation of the SAR requires knowledge of transmission from a primary case to susceptible contacts in their transmission unit. To confirm an infection in a contact as a transmission event, acquisition of infection from the community and contact-to-contact transmission within a transmission unit need to be ruled out. Including community acquisition or contact-to-contact transmission events generally biases the calculation of the SAR upwards.[76] The direction and magnitude of the bias in the VE against the SAR will depend on the extent to which contacts of vaccinated vs. unvaccinated primary cases acquire infection from the community and from other transmission unit members.

### *3.5.1 Community-acquired infections in susceptible contacts*

Figure 1[d] illustrates the issue with community-acquired infections: without careful attention, we may mistakenly classify infections diagnosed after the index case as transmission events when in fact the infections are community-acquired. One way to account for community acquisition is through modeling, with assumptions around the latent and infectious periods for both vaccinated and unvaccinated individuals,[76–79] as applied to SARS-CoV-2 by Prunas and colleagues.[22] An alternative approach is to leverage additional data to distinguish between community-acquired and transmission events. One source of information is the timing of diagnosis; for example, Gier and colleagues [20] and Harris and colleagues [21] include only infected contacts who developed symptoms within 2-14 days of the index case. Viral genetic sequence data may also assist in ruling out potential transmission events.[80,81] High-quality SARS-CoV-2 genome sequencing data and epidemiological data have been used during outbreak investigations.[82,83] Studies of infection clustering use viral genetic testing to construct and infer phylogenetic trees, including studies of SARS-CoV-2 transmission among university students [84] and U.S. Marine recruits.[85] Finally, epidemiologic data can assist in restricting attention to possible or likely transmission events based on close contact. For example, Sikkens and colleagues used test results and behavioral data to rule out cases of healthcare worker to patient transmission.[86]

The choices of transmission unit and study duration can assist in distinguishing between community acquisition events and transmission events. For household studies, the within-household SAR for SARS-CoV-2 during Omicron has been estimated at 50-70%.[24] In contrast, the likelihood of commuity-acquired infection over a short time period, even in the context of a local outbreak, is generally smaller. For example, even during the peak of the Omicron epidemic in New York City with nearly 700,000 recorded incident infections in January 2022, about 8% of residents were diagnosed with infection.[87] In general, the greater the ability of the study to restrict attention to tightly-interacting transmission units over the duration of infectiousness for the primary case, the better the chances are that commuity acquisition events can be excluded.

### *3.5.2 Contact-to-contact transmission within the same transmission unit*

A second scenario that needs to be ruled out is when a susceptible contact acquired a SARS-CoV-2 infection from a different contact within the transmission unit, other than the primary case (Figure 1[e]). Given that a transmission unit is selected as a group of individuals who have frequent, close physical contact, it is difficult to rule out entirely the possibility of contact-to-contact transmission even with timing of diagnosis, viral genetic data, and behavioral data. Nevertheless, failing to account for contact-to-contact transmission may introduce bias in the estimation of the SARS-CoV-2 SAR, and thus bias in the estimation of vaccine efficacy against the SAR.[76] One strategy is a model-based approach such as that explored by Prunas and colleagues,[22] where infection times and infection durations are simulated to infer pairwise transmission risk.[77] These strategies may help limit misclassification of transmission events.

The SARS-CoV-2 viral load in the index case and infected contact, which is a likely surrogate for transmission potential,[88–91] may assist in excluding contact-to-contact transmission events. Propectively-designed studies have the important merit that they can capture full viral load curves, thus greatly enhancing accurate timing of acquisition in the contact and of infectiousness of the index. However, retrospective studies seldom capture viral load in indexes or contacts, as this is not a routine measurement in clinical practice. Some qualitative PCR assays report cycle threshold (Ct) as a surrogate of viral load; however, there is considerable variability across assays and platforms and specimen types in terms of the reference range and reliability of this measure.[89,90] Even when measured, retrospective studies only have the Ct measurement at a single time point (at infection diagnosis) which greatly limits the utility of the measure for classifying transmission events.

# 4 Discussion

More than two years into the pandemic, even though COVID-19 vaccines have been shown highly effective at reducing COVID-19 disease, hospitalization, and death, the ongoing SARS-CoV-2 pandemic remains a major global public health challenge. As more transmissible variants have arisen, in the context of high rates of vaccine hesitancy in some populations and scarce access to vaccine in other populations, we need definitive evidence on the overall ability of COVID-19 vaccines to reduce onward transmission of infection.

While there remain arguments for randomized trial designs in settings without access to COVID-19 vaccines [92,93] and for randomized rollout of vaccines in settings where vaccines are starting to be distributed,[33,36,94,95] future studies with randomized designs and clinical endpoints are likely to be few. Furthermore, because capturing transmission events through prospective and frequent testing is resource-intensive, future studies of COVID-19 vaccine effects on transmission are likely to be observational.

Observational designs that follow potential transmission units prospectively will provide the most rigorous answers. This may be accomplished by prospective enrollment and testing of entire transmission units, prior to any infection diagnosis, and follow-up of all members over a long period spanning an outbreak, e.g., months.[33,96,97] Cohen and colleagues implemented such a prospective design, initiated in the context of seasonal influenza [98] and pivoted to study SARS-CoV-2 at the onset of the pandemic,[99] where individuals within household transmission units performed twice-weekly nasal swabs to permit PCR testing regardless of symptoms for thirteen months. Alternatively, a 'case-ascertained' approach enrolls transmission units with an incident infection, all of whom are tested for a short time period covering the likely infectious period of the index case.[96,100] Clifford and colleagues [101] implemented a case-ascertained approach where household contacts of index cases in the UK were tested for SARS-CoV-2 on days 1, 3, and 7 after enrollment. Prospective testing of contacts will provide the most accurate capture and classification of transmission events but is resource intensive.

Retrospective designs for studies of SARS-CoV-2 transmission include record-linkage for individuals by address, or contact-tracing for confirmed COVID-19 cases. While these databases are rich sources of data, they encounter several limitations due to their retrospective nature and the fact that the associated data are collected for non-research purposes. Infrequent and irregular testing is likely to miss capturing infections, with longer-duration and symptomatic infections more likely captured in both index cases and their susceptible contacts.[29] Misclassification of the primary cases and secondary transmission events is also possible, perhaps even likely. There is likely to be incomplete testing of both indexes and contacts, and a limited understanding of drivers of testing utilization. Furthermore, even with correct identification of primary cases, there may be bias due to the inclusion of community-acquired infections and contact-to-contact transmission. Finally, factors that are associated with both risk of acquisition or transmission of SARS-CoV-2 and with uptake and timing of COVID-19 vaccination, or with both receipt of SARS-CoV-2 tests and vaccination, are potential cofounders. While methods exist to adjust for confounding due to measured factors, a larger concern is bias due to unmeasured factors, since variables captured in retrospective databases are likely not comprehensive. Specifically, behavioral factors that influence pathogen exposure and transmission, uptake of vaccination, and testing strategies, are likely unmeasured.

**Considerations for Statistical Analysis.** Improved statistical analyses may help in addressing some of the challenges encountered in retrospective studies of vaccine efficacy against the SAR for SARS-CoV-2 and other respiratory pathogens. The fundamental challenge encountered for both record-linked and contact-tracing databases is informatively missing data: missed primary cases and transmission events, and misclassified transmission events. Given additional information or knowledge on the mechanism of missingness, it is possible to employ statistical corrections. For example, missed primary cases and transmission events can be corrected by adjusting for the rates at which asymptomatic infections of vaccinated and unvaccinated individuals are detected, if known. Given more complex and varied drivers of missingness, statistical models may be leveraged to model the probability of missing data as a function of measured variables, and inverse-probability-weighted (IPW), or more efficient 'augmented' IPW analyses, may be performed to attempt to account for the informative missingness.[102,103] Latent variable analyses which allow for the modeling of un-observed variables, may also be informative here.[104]

**Recommendations for Study Design.** In terms of study design, we recommend leveraging studies in populations that have controlled diagnostic testing prompts over the duration of study, such as healthcare workers or university students, as this can mitigate the effects of outcome-driven testing. We recommend that authors perform sensitivity analyses around factors related to the testing program to describe potential directions of bias in vaccine efficacy estimates, as in our Figure 1, and to provide careful interpretation of the vaccine efficacy estimand that was estimated given the testing program. It is also important, as many of the retrospective studies have done, to carefully examine infections in susceptible contacts to include only plausible transmission events, rather than all infections within the unit. Using epidemiological data combined with phylogenetic analyses can supplement this effort.

**Recommendations for Reporting Results.** For reporting results, we recommend that authors follow the STROBE guidelines for observational studies,[105,106] and specifically the RECORD guidelines for observational routinely collected health data.[107] In this particular context, we recommend that authors describe the testing program in place during the study including drivers of testing utilization and the extent of testing completion (e.g., percent of household members receiving a test within 1 week of the index case, see Lyngse and colleagues Figure 1[a][23]), and summarize how these vary among vaccinated vs. unvaccinated individuals and contacts. We recommend that the authors provide explicit details about how transmission events were ascertained, the expected level of community transmission that occurred during the study, and how the authors accounted for contact-to-contact transmission. In addition to describing which variables were used to control for confounding, we recommend including a discussion of which unmeasured variables the authors believe may contribute to residual confounding given the specific study population, time, and local epidemic dynamics. These strategies will assist readers in interpreting the vaccine efficacy parameters in the proper context and gauging the reliability and generalizability of the study results.

.

**Table 1.** Statistical designs of example studies of COVID-19 vaccine efficacy against SARS-CoV-2 transmission. Studies were selected as examples to illustrate key elements of the retrospective designs discussed. This is not an exhaustive list of retrospective studies of vaccine efficacy against the SARS-CoV-2 SAR.

| Reference | Design | Testing strategy | Transmission Unit and Susceptible Contacts | Strategy to exclude community transmission | Strategy to restrict to transmissions from primary subject only |
|---|---|---|---|---|---|
| Eyre et al 2022 | Analysis of contact tracing database | Event-prompted testing for primary subjects.<br>Only close contacts that received a test were included. | Relative risk of infection in close contacts that received a PCR test within 1-10 days of the index case, comparing index cases that were un-vaccinated, partially vaccinated (1 dose of AstraZeneca or Pfizer), and fully vaccinated (2 doses of AstraZeneca or Pfizer), controlling for vaccination status of contact. | *N/A: No pre-defined transmission unit.* | Included only test results from contacts tested within 1-10 days of index case |
| Gier et al. 2021 | Analysis of contact tracing database<br>Person can be both a contact and an index | Event-prompted testing for primary subjects.<br>Encouraged contacts of index case to get tested as soon as possible, and on the fifth day after last exposure. | Relative risk of infection in close contacts of un-vaccinated adults compared to adults with confirmed infection that have received partial or full vaccination (Pfizer, Moderna, or Janssen) in the Netherlands.<br>Both pooled and stratified analyses by vaccination status of the contact. | *N/A: No pre-defined transmission unit.* | Excluded indexes where their most likely source of infection was in the home<br>Included only test results from contacts tested within 1-14 days of index case<br>Required quarantining of contacts after exposure |
| Harris et al. 2021 | Analysis of healthcare databases with linked records for households | Event-prompted testing for primary subject and contacts. | Relative risk of infection in un-vaccinated household contacts of individuals who have received Pfizer or AstraZeneca at least three weeks prior to testing positive for infection to compared to un-vaccinated individuals who also test positive. | Only include contact infections that occurred within 2-14 days of index | Exclude households with more than one person testing positive within two days |
| Lyngse et al. 2021 | Analysis of healthcare databases with linked records for households | Event-prompted testing for primary subjects.<br>Substantial effort put into contact tracing. | Relative risk of infection in household contacts (controlling for vaccination status) of fully vaccinated [14+ days after] and boosted [7+ days after] individuals who test positive compared to partially or un-vaccinated individuals who also test positive.<br>*Paper specifically focuses on comparing this relative risk between households with primary Omicron infection versus primary Delta infection.* | Only include contact infections that occurred within 1-7 days of index. | Exclude households with more than 1 person testing positive on same day. |
| Prunas et al. 2021 | Analysis of healthcare databases with linked records for households<br>Does not assign an index case, models acquisition and transmission events within households over time by imputing infection times | Event-prompted testing for primary subject and contacts. | Relative risk of infection in household contacts of vaccinated subjects compared to un-vaccinated subjects, controlling for vaccination status of contact. | Allow for baseline risk of community-acquired infection in modeling | Imputation of infection times allows for *within-household sources* of infections that are not the primary subject in modeling |

# Appendix A: Derivations for potential impact of event-prompted or infrequent testing on estimation of vaccine efficacy

To illustrate the potential impact of event-prompted or infrequent testing on estimation of VE against transmission, we compare the "target" estimand — the expectation of the statistical estimate of vaccine efficacy against the SARS-CoV-2 SAR — with the "actual" estimands under imperfect testing. We assume a range of scenarios around SARS-CoV-2 infection and transmission that are intended to illustrate the issues of symptom-prompted testing and infrequent testing in isolation; in practice there may be several issues at play. We describe these scenarios below.

In the reference scenario, we assume that we have a prospectively defined group so that we correctly identify the primary case if they are tested, and any active infection is always identified if they are tested. For simplicity, all contacts are un-vaccinated and if infected the symptom presentation and duration of infection is unrelated to attributes of the primary case or transmission unit. Therefore, symptom-prompted, or infrequent testing of contacts does not change the estimand. Reflecting the high efficacy of COVID-19 vaccines against symptomatic infection, we fix vaccine efficacy against symptomatic infection at 0.9. Vaccine efficacy against SARS-CoV-2 infection is set at 0.5, so that the proportion of symptomatic infections in the vaccinated is 20% of that in the un-vaccinated. Given infection, the probability of symptoms among the un-vaccinated is 0.5, and the reduction in SAR for asymptomatic vs. symptomatic infections, and the overall VE against the SAR (the target estimand) are varied. Under infrequent testing, we assume that the duration of infection is uniformly distributed over 14 days, with a mean of 14 days for the un-vaccinated, and 8 days for the vaccinated. We vary the frequency of testing from once every 1 day to once every 14 days.

## Symptom-Prompted Testing

Let $\mathrm{T}$ be an indicator for transmission, $\mathrm{S}$ be an indicator for being symptomatic, and $\mathrm{V}$ be an indicator for vaccinated. Let $\mu$ represent the ratio of transmission probability between those vaccinated and un-vaccinated with $0 \leq \mu \leq 1$, i.e.,

$$P(T = 1|V = 1) = \mu P(T = 1|V = 0).$$

Vaccine efficacy is then defined as $1 - \mu$. We can rewrite each of the probabilities in terms of symptom development:

$$\begin{aligned} P(T = 1|V = 0) &= P(T = 1|S = 1, V = 0)P(S = 1|V = 0) + P(T = 1|S = 0, V = 0)P(S = 0|V = 0) \\ P(T = 1|V = 1) &= P(T = 1|S = 1, V = 1)P(S = 1|V = 1) + P(T = 1|S = 0, V = 1)P(S = 0|V = 1). \end{aligned}$$

We now define the following five parameters, and we will rewrite the vaccine efficacy parameter $1 - \mu$ as a function of these parameters.

$$\begin{aligned}
\lambda \quad &:= \frac{P(S=1|V=1)}{P(S=1|V=0)} \\
\delta \quad &:= \frac{P(T=1|S=0,V=v)}{P(T=1|S=1,V=v)} \\
\nu \quad &:= \frac{P(T=1|S=s,V=1)}{P(T=1|S=s,V=0)} \\
\rho \quad &:= P(S=1|V=0) \\
\tau \quad &:= P(T=1|S=1,V=0).
\end{aligned}$$

That is, $\lambda$ represents the reduction in symptoms given vaccination, $\delta$ represents the reduction in transmission potential for asymptomatic individuals, holding vaccine status constant, and $\nu$ represents the reduction in transmission potential for vaccinated individuals, holding symptom presence constant. The probability of symptoms in a vaccinated person is $\rho$, and the probability of transmission for a symptomatic, un-vaccinated person is $\tau$. Our goal is to solve for $\mu$ given fixed values of these quantities. Replacing the earlier probabilities with these relationships,

$$\begin{aligned}
P(T=1|V=0) \quad &= P(T=1|S=1,V=0)P(S=1|V=0) + P(T=1|S=0,V=0)P(S=0|V=0) \\
&= \rho P(T=1|S=1,V=0) + (1-\rho)P(T=1|S=0,V=0) \\
&= \rho P(T=1|S=1,V=0) + (1-\rho)\delta P(T=1|S=1,V=0) \\
&= \tau\{\rho + (1-\rho)\delta\} \\
P(T=1|V=1) \quad &= P(T=1|S=1,V=1)P(S=1|V=1) + P(T=1|S=0,V=1)P(S=0|V=1) \\
&= P(T=1|S=1,V=1)P(S=1|V=1) + P(T=1|S=0,V=1)\big(1 - P(S=1|V=1)\big) \\
&= P(T=1|S=1,V=1)\lambda\rho + P(T=1|S=0,V=1)(1-\lambda\rho) \\
&= P(T=1|S=1,V=1)\lambda\rho + P(T=1|S=1,V=1)\delta(1-\lambda\rho) \\
&= P(T=1|S=1,V=0)\nu\lambda\rho + P(T=1|S=1,V=0)\delta(1-\lambda\rho)\nu \\
&= \tau\nu\{\lambda\rho + \delta(1-\lambda\rho)\}.
\end{aligned}$$

Thus, using the relationship $P(T=1|V=1) = \mu P(T=1|V=0)$, we can find the vaccine efficacy $\mu$:

$$\begin{aligned}
P(T=1|V=1) - \mu P(T=1|V=0) \quad &= \tau\nu\{\lambda\rho + \delta(1-\lambda\rho)\} - \mu\tau\{\rho + (1-\rho)\delta\} \equiv 0 \\
&\Rightarrow \nu\{\lambda\rho + \delta(1-\lambda\rho)\} = \mu\{\rho + (1-\rho)\delta\} \\
&\Rightarrow \mu = \frac{\nu\{\lambda\rho + \delta(1-\lambda\rho)\}}{\rho + \delta(1-\rho)}
\end{aligned}$$

With only symptomatic sampling of the index cases, we would instead have

$$P(T=1|S=1,V=1) - \tilde{\mu}P(T=1|S=1,V=0) = \tau\nu - \tau\tilde{\mu}$$

so, in this case, $\tilde{\mu} = \nu$, exactly the reduction in transmission given vaccination, holding symptom status constant. If vaccination has no effect on development of symptoms, $\lambda = 1$, and we can see that (1) reduces to $\mu = \nu$. Therefore, the parameter being estimated when we only sample symptomatic index cases is the reduction in transmission with vaccination, only among cases that become symptomatic.

## Infrequent Testing

Let $\mathrm{T}$ be an indicator for transmission, $\mathrm{N}$ be a positive-valued random variable indicating duration of infection, and $\mathrm{V}$ be an indicator for vaccinated. As before, let $\mu$ represent the ratio of transmission probability between those vaccinated and un-vaccinated with $0 \leq \mu \leq 1$, i.e.,

$$\mathrm{P}(\mathrm{T}=1|\mathrm{V}=1) = \mu \mathrm{P}(\mathrm{T}=1|\mathrm{V}=0).$$

We can rewrite each of the probabilities in terms of duration of infectiousness:

$$\begin{aligned} \mathrm{P}(\mathrm{T}=1|\mathrm{V}=0) \quad &= \int \mathrm{P}(\mathrm{T}=1|\mathrm{N}=\mathrm{n},\mathrm{V}=0)\mathrm{dP}_{\mathrm{N|V=1}}(\mathrm{n}) \\ \mathrm{P}(\mathrm{T}=1|\mathrm{V}=1) \quad &= \int \mathrm{P}(\mathrm{T}=1|\mathrm{N}=\mathrm{n},\mathrm{V}=1)\mathrm{dP}_{\mathrm{N|V=0}}(\mathrm{n}). \end{aligned}$$

Let $\mathrm{P}_{\mathrm{N|V=0}}(\mathrm{n})$ and $\mathrm{P}_{\mathrm{N|V=1}}(\mathrm{n})$ be the distribution function of infection duration for infected people who are un-vaccinated and vaccinated, respectively. Let $\tau_0$ be the daily risk of transmission to a contact among the un-vaccinated and $\tau_1 = \tau_0 \nu$, $\nu \leq 1$ be the daily risk of transmission to a contact among the vaccinated. Define $\nu = \tau_1/\tau_0 \leq 1$. We assume that there is a constant risk of infection across the duration of infection ($\mathrm{P}(\mathrm{T}=1|\mathrm{N}=\mathrm{n},\mathrm{V}=\mathrm{v}) = 1 - \exp(-\mathrm{n}\tau_\mathrm{v}) \approx \mathrm{n}\tau_\mathrm{v}$ for small $\tau_\mathrm{v}$). Then,

$$\begin{aligned} \mathrm{P}(\mathrm{T}=1|\mathrm{V}=0) \quad &= \int \mathrm{n}\tau_0 \mathrm{dP}_{\mathrm{N|V=0}}(\mathrm{n}) \\ \mathrm{P}(\mathrm{T}=1|\mathrm{V}=1) \quad &= \int \mathrm{n}\tau_1 \mathrm{dP}_{\mathrm{N|V=1}}(\mathrm{n}). \end{aligned}$$

If we assume that $\mathrm{P}_{\mathrm{N|V=0}}(\mathrm{n}) = \mathrm{Unif}(\rho_0 - \mathrm{c}, \rho_0 + \mathrm{c})$, $\mathrm{P}_{\mathrm{N|V=1}}(\mathrm{n}) = \mathrm{Unif}(\rho_1 - \mathrm{c}, \rho_1 + \mathrm{c})$, and with $\lambda = \rho_1/\rho_0 \leq 1$ (and in particular, to have a sensible minimum duration of infection among the vaccinated [i.e., $\rho_1 - \mathrm{c}$], we need to have $\lambda > \mathrm{c}/\rho_0$), then we have,

$$\begin{aligned} \mathrm{P}(\mathrm{T}=1|\mathrm{V}=0) \quad &= \tau_0 \mathbb{E}_{\mathrm{N|V=0}}\{\mathrm{N}\} \\ &= \tau_0 \rho_0 \\ \mathrm{P}(\mathrm{T}=1|\mathrm{V}=1) \quad &= \tau_1 \mathbb{E}_{\mathrm{N|V=1}}\{\mathrm{N}\} \\ &= \tau_1 \rho_1 \\ &= \tau_0 \rho_0 \lambda \nu. \end{aligned}$$

Therefore, we have $\mu = \lambda\nu$.

Let $\mathrm{S}$ be an indicator of the index case being included in the sample. Let $\mathrm{k}$ be the window of time during which an infection may be sampled (e.g., $\mathrm{k} = 14$ means there is a two-week period when someone is infected during which a single test can be taken). In this testing scenario we consider $\mathrm{P}(\mathrm{S}=1|\mathrm{N}=\mathrm{n}) = \max(\mathrm{n}/\mathrm{k}, 1)$, i.e., the probability of being sampled is proportional to the duration of infection, and it is dependent on no other variables. For fixed vaccine status (where we note that with $\mathrm{P}_{\mathrm{N|V=v}}$ defined above, $\mathrm{Var}_{\mathrm{N|V=v}}\{\mathrm{N}\} = \mathrm{c}^2/3$), we have:

$$
\begin{aligned}
P(T=1|V=v,S=1) &= \int P(T=1|V=v,S=1,N=n)P(N=n,|V=v,S=1)dn \\
&= \int P(T=1|V=v,N=n)P(N=n|V=v,S=1)dn \quad \text{since } T \perp S|N \\
&= \int n\tau_v \frac{P(S=1|N=n,V=v)P(N=n|V=v)}{P(S=1,V=v)} dn \quad \text{by Bayes formula} \\
&= \tau_v \int n \frac{P(S=1|N=n)}{\int P(S=1|N=n,V=v)dP_{N|V=v}(n)} dP_{N|V=v}(n) \quad \text{since } S \perp V|T \\
&= \tau_v \int n \frac{P(S=1|N=n)}{\int P(S=1|N=n)dP_{N|V=v}(n)} dP_{N|V=v}(n) \quad \text{since } S \perp V|T
\end{aligned}
$$

We first consider the case where $k > \rho + c$, i.e., no one has a probability of testing greater than 1. In that case, we can have only one integral:

$$
\begin{aligned}
&= \tau_v \int n\{n/k\} \frac{1}{P(S=1|V=v)} dP_{N|V=v} \\
&= \tau_v \int n\{n/k\} \frac{1}{\int P(S=1|N=n,V=v)dP_{N|V=v}} dP_{N|V=v} \\
&= \tau_v \int n^2(1/k) \frac{1}{\int (n/k)dP_{N|V=v}} dP_{N|V=v} \\
&= \tau_v \mathbb{E}_{N|V=v}\{N^2\}/\mathbb{E}_{N|V=v}\{N\} \\
&= \frac{\tau_v}{\rho_v}\{\mathbb{E}_{N|V=v}\{N\}^2 + \mathrm{Var}_{N|V=v}\{N\}\} \\
&= \frac{\tau_v}{\rho_v}\{\rho_v^2 + (1/3)c^2\}.
\end{aligned}
$$

Interestingly, the parameter for the testing window size $k$ drops out of the expression. Thus, the inferred parameter for $\mu$, $\tilde{\mu}$ – when $k > \rho_0$ (which implies $k > \rho_1$ as well) is given by

$$
\begin{aligned}
\tilde{\mu} &= \frac{P(T=1|V=1,S=1)}{P(T=1|V=0,S=1)} \\
&= \frac{\tau_1/\tau_0}{\rho_1/\rho_0}\{\frac{\rho_1^2 + (1/3)c^2}{\rho_0^2 + (1/3)c^2}\} \\
&= \frac{\nu}{\lambda}\{\frac{\lambda^2\rho_0^2 + (1/3)c^2}{\rho_0^2 + (1/3)c^2}\}.
\end{aligned}
$$

Now consider the setting where $k \in [\rho_v - c, \rho_v + c]$. In this case people with duration $n \geq k$ will be sampled with probability 1. The denominator in the probability is $\int P(S=1|N=n)dP_{N|V=v}(n)$, which can be evaluated as (we use $dP$ for simplicity):

$$\begin{aligned}
\int P(S=1|N=n)dP_{N|V=v}(n) &= \int P(S=1|N=n)1(n\le k)dP(n) + \int P(S=1|N=n)1(n>k)dP(n) \\
&= \int_{\rho_v-c}^{k} (n/k)dP(n) + \int_{k}^{\rho_v+c} d\,P(n) \\
&= \frac{1}{2ck}\{\int_{\rho_v-c}^{k}(n)dn\} + \frac{1}{2c}\{\int_{k}^{\rho_v+c} d\,n\} \\
&= \frac{1}{2ck}\{\frac{k^2-(\rho_v-c)^2}{2}\} + \frac{1}{2c}\{(\rho_v+c)-k\} \\
&= \frac{1}{4ck}\{k^2-(\rho_v-c)^2+2k(\rho_v+c)-2k^2\} \\
&= \frac{1}{4ck}\{2k(\rho_v+c)-(\rho_v-c)^2-k^2\} =: S_k.
\end{aligned}$$

Making sure that this still equals $\rho_v/k$, we have when $k=(\rho_v+c)$ that:

$$\begin{aligned}
\int P(S=1|N=n)dP_{N|V=v}(n) &= \frac{1}{4c(\rho_v+c)}\{2(\rho_v+c)^2-(\rho_v-c)^2-(\rho_v+c)^2\} \\
&= \frac{1}{4c(\rho_v+c)}\{(\rho_v+c)^2-(\rho_v-c)^2\} \\
&= \frac{\rho_v^2+2\rho_v c+c^2-\rho_v^2+2\rho_v c-c^2}{4c(\rho_v+c)} \\
&= \frac{4\rho_v c}{4c(\rho_v+c)} \\
&= \frac{\rho_v}{\rho_v+c}.
\end{aligned}$$

So, this is equivalent to what we got for the denominator expression in the first derivation when $n<k$. We now derive the full expression for $P(T=1|V=v,S=1)$ and check to make sure that we get the same thing as in (2) when $k=(\rho_v+c)$.

Putting this together with the numerator we have:

$$\begin{aligned}
P(T=1|V=v,S=1) &= \frac{\tau_v}{S_k}\{\int n(n/k)1(n\le k)dP(n) + \int n1(n>k)dP(n)\} \\
&= \frac{\tau_v}{S_k}\{\frac{1}{k}\int_{\rho_v-c}^{k} n^2\,dP(n) + \int_{k}^{\rho_v+c} n\,dP(n)\} \\
&= \frac{\tau_v}{S_k}\{\frac{1}{k}\int_{\rho_v-c}^{k} n^2\,dP(n) + \int_{k}^{\rho_v+c} n\,dP(n)\} \\
&= \frac{\tau_v}{S_k 2c}\{\frac{1}{k}\int_{\rho_v-c}^{k} n^2\,dn + \int_{k}^{\rho_v+c} n\,dn\} \\
&= \frac{\tau_v}{S_k 2c}\{\frac{1}{3k}\{k^3-(\rho_v-c)^3\} + \frac{(\rho_v+c)^2-k^2}{2}\} \\
&= \frac{\tau_v}{12ckS_k}\{2k^3-2(\rho_v-c)^3+3k(\rho_v+c)^2-3k^3\} \\
&= \frac{\tau_v}{12ckS_k}\{3k(\rho_v+c)^2-k^3-2(\rho_v-c)^3\}.
\end{aligned}$$

When $k=(\rho_v+c)$,

$$
\begin{aligned}
P(T=1|V=v,S=1) &= \frac{\tau_v(\rho_v+c)}{12c\rho_v(\rho_v+c)}\{3(\rho_v+c)^3-(\rho_v+c)^3-2(\rho_v-c)^3\} \\
&= \frac{\tau_v}{12c\rho_v}\{2(\rho_v+c)^3-2(\rho_v-c)^3\} \\
&= \frac{\tau_v}{6c\rho_v}\{(\rho_v+c)^3-(\rho_v-c)^3\} \\
&= \frac{\tau_v}{6c\rho_v}\{\rho_v^3+3\rho_v^2c+3\rho_v c^2+c^3-\rho_v^3+3\rho_v^2c-3\rho_v c^2+c^3\} \\
&= \frac{\tau_v}{6c\rho_v}\{6\rho_v^2c+2c^3\} \\
&= \frac{\tau_v}{\rho_v}\{\rho_v^2+(1/3)c^2\}.
\end{aligned}
$$

A third scenario is where $k < \rho_v - c$, such that all infections are captured with probability 1. In this case, $\int P(S=1|N=n)dP(n)=1$, and $P(T=1|V=v,S=1)=\tau_v\int ndP(n)=\tau_v\rho_v$. Therefore,

$$
\begin{aligned}
\tilde{\mu} &= \frac{P(T=1|V=1,S=1)}{P(T=1|V=0,S=1)} \\
&= \frac{\tilde{\mu}_1}{\tilde{\mu}_0}
\end{aligned}
$$

where

$$
\tilde{\mu}_v = \tau_v\{\rho_v 1(k<\rho_v-c)+\{\rho_v+(1/3\rho_v)c^2\}1(\rho_v-c\le k\le\rho_v+c)+\frac{3k(\rho_v+c)^2-k^3-2(\rho_v-c)^3}{12ckS_k}1(k>\rho_v+c)\}.
$$

Importantly, $\tilde{\mu}_1$ and $\tilde{\mu}_0$ may differ with respect to which of the indicator functions is activated, since $\rho_1$ will be different than $\rho_0$ when $\lambda < 1$. Notice that if $k >$ max$(\rho_0, \rho_1)$, then the ratio of $\tilde{\mu}_1/\tilde{\mu}_0$ does not depend on $k$. There is still a difference between $\tilde{\mu}$ and $\mu$, but it does not change with the interval $k$. We can see this relationship in Figure A1.

**Figure A1.** Relationship between the inferred vaccine efficacy parameter $1 - \tilde{\mu}$ and $k$, the testing interval. Colors indicate the true vaccine efficacy, which is also obtained by setting the testing interval to 1 day.

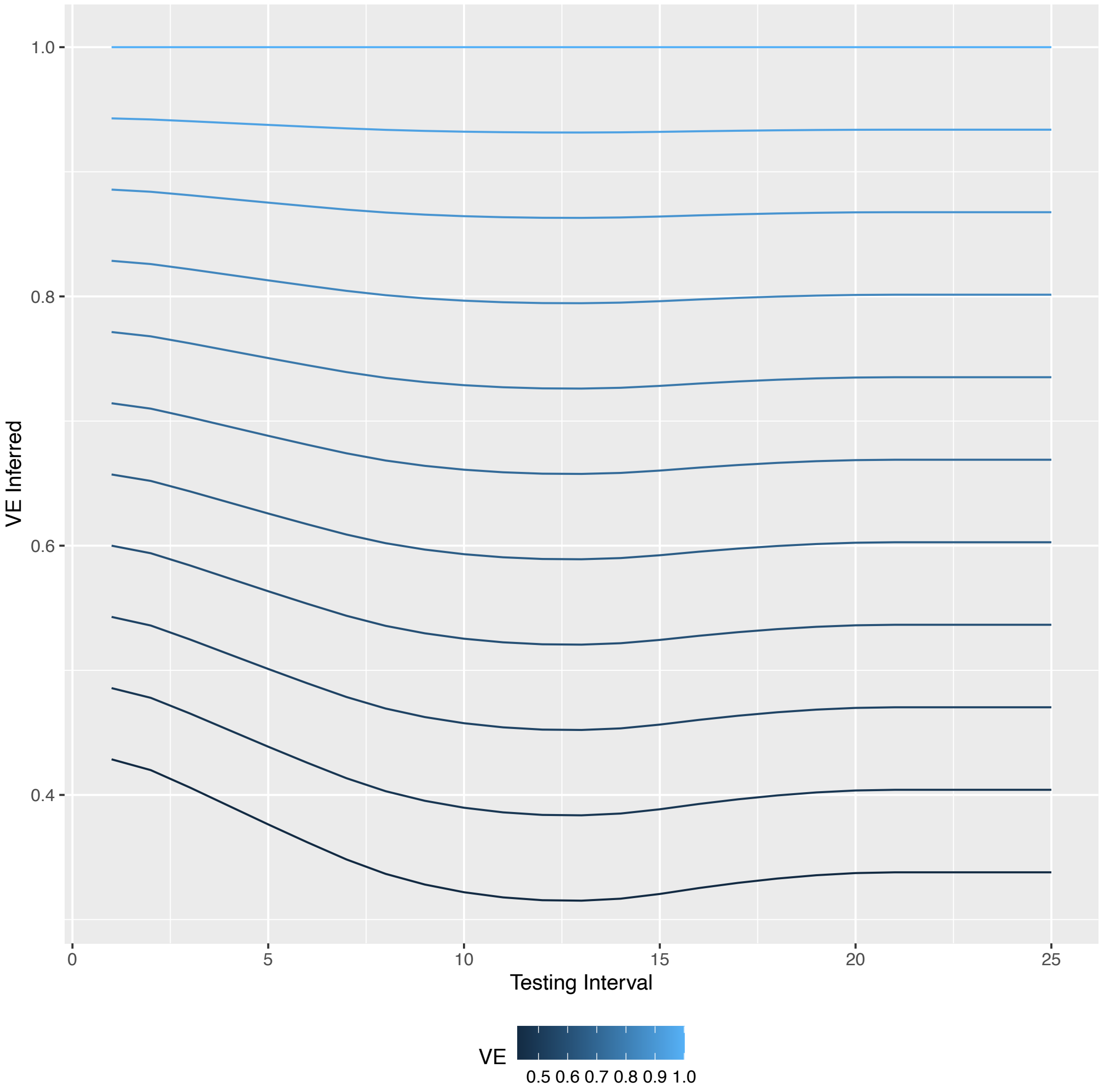